# Atomic manipulation of the gap in $Bi_2Sr_2CaCu_2O_{8+x}$

F. Massee[1*], Y. K. Huang[2], M. Aprili[1].

[1]Laboratoire de Physique des Solides (CNRS UMR 8502), Bâtiment 510, Université Paris-Sud/Université Paris-Saclay, 91405 Orsay, France.

[2]Institute of Physics, University of Amsterdam, 1098XH Amsterdam, The Netherlands.

*Correspondence to: freek.massee@u-psud.fr.

**Single atom manipulation within doped correlated electron systems would be highly beneficial to disentangle the influence of dopants, structural defects and crystallographic characteristics on their local electronic states. Unfortunately, their high diffusion barrier prevents conventional manipulation techniques. Here, we demonstrate the possibility to reversibly manipulate select sites in the optimally doped high temperature superconductor $Bi_2Sr_2CaCu_2O_{8+x}$ using the local electric field of the tip. We show that upon shifting individual Bi atoms at the surface, the spectral gap associated with superconductivity is seen to reversibly change by as much as 15 meV (~5% of the total gap size). Our toy model that captures all observed characteristics suggests the field induces lateral movement of point-like objects that create a local pairing potential in the $CuO_2$ plane.**

One of the challenges in the study of high temperature superconductivity, and its relation to various charge- and spin ordered phases in the cuprates (*1*), is their intrinsic inhomogeneous nature. This is exemplified by the archetypal system $Bi_2Sr_2CaCu_2O_{8+x}$ (Bi2212), whose complicated crystal structure includes an incommensurate structural super-modulation (*2*) and interstitial oxygen dopant atoms. Particularly striking is the large variation in spectral gap size over nanometre-scale distances (*3-5*), which has been shown to reflect local variations in $T_c$ (*6, 7*) and has been correlated to both oxygen dopants (*8, 9*) and structural inhomogeneity (*10*). These correlations, however, typically involve averages over a large number of sites. In order to directly probe the relation between dopants, structural defects and crystallographic characteristics on the local electronic states at the atomic scale, direct non-invasive control of the dopant positions and the inhomogeneous crystal structure they inhabit is therefore desirable. Unfortunately, common techniques for single atom manipulation involving short-range forces between the tip and the atom (*11-13*) and/or vibrational excitation using the tunnelling current (*14-16*) are unsuited to controllably manipulate atoms in single crystal cuprate materials as the dopant atoms are buried under the surface and the diffusion barrier of the surface atoms themselves is too high. Alternatively, the electric field can be used to manipulate a surface (*17-19*), although since the field profile depends on the size of the tip apex, this process is in practice difficult to control - unless specific atoms are more sensitive to the field that others due to their charge or local environment. We discovered that this is exactly the case in Bi2212, where we find two local environments that are more readily influenced by the electric field than the rest of the system.

Figure 1a shows a schematic of the two environments that allow for field induced atom manipulation: recently discovered weakly coupled oxygen dopants (*20*) and surface Bi atoms on the crest of the super-modulation. To manipulate oxygen dopants, which we identify from their resonance at negative energies (*8, 9, 20*), we position our tip above a dopant and slowly increase the sample bias voltage, $V_s$. Upon reaching ~800 mV at a tunnel current of ~100 pA, we observe jumps in the current (see Fig. 1b) that we can freeze-in by switching back to low bias voltage. For different currents thus selected, the resonance of the dopant at negative energies is seen to shift in energy (see Fig. 1c). Given that the energy of the resonance is closely related to the depth of the dopant (*21*), this strongly suggests that the electric field of the tip pushes the negatively charged dopant away from the tip, towards the CuO$_2$ plane. Since the spatial extent of the highest electric field is determined by the size of the tip apex, which is typically a few tens of nanometres in diameter, in principle several hundred dopants can be affected. However, unlike for a homogeneous system where the threshold field for manipulation is identical for all field-affected entities (*19*), the existence of numerous local dopant environments in Bi2212 (*20*) leads to a range of threshold fields. By positioning the tip directly above the dopant with the lowest of these threshold fields, selective manipulation is possible. For fields exceeding the lowest threshold field, however, two processes may occur: 1) the dopant with the lowest threshold field is manipulated more vigorously and 2) the threshold field for other dopants is reached. Indeed, upon increasing the manipulation voltage, we observe changes on multiple dopant locations (see Supplementary Information section I). Additionally, for $V_s \geq 1.2$ V, surface Bi atoms on the crest of the super-modulation - which are likely more mobile than others due to the lattice distortion of the super-modulation itself - are manipulated, see Fig. 1d. As expected for electric field induced manipulations, all changes we observe are confined to a roughly circular area with a diameter of a few tens of nanometres around the location of the tip during the application of the manipulation voltage (see Fig. S4).

With the ability to locally rearrange the dopant and Bi surface atoms, we are now in a position to study how these structural modifications affect the low energy electronic states. To this end, we record low energy differential conductance maps and extract a map of the peak-to-peak gap, $\Delta_i(\mathbf{r})$, using identical settings before and after applying an electric field at a sample bias of 1.3-1.5 V and ~100 pA. This field is sufficient to manipulate on the order of ten dopants and Bi surface atoms. Fig. 2 shows the result for one treatment at $V_s = 1.5$ V, three subsequent treatments are shown in Supplementary Information section IV. Whereas the topographies (Fig. 2a-b) and the gap maps (Fig. 2c-d) are mostly identical, which excludes any change of the tip itself, select locations show significant changes, with gaps increasing and decreasing by over 10 meV. The two spectra in Fig. 2e, which are taken at the same location before and after the electric field induced manipulation, highlight the significance of these changes. Interestingly, the spectra are predominantly affected at the peak energies, whereas the low energy states, $|E| < 20$ meV, are hardly, if at all modified, highlighting the insensitivity of the latter to disorder (*22*), see also Supplementary Information section III. Lastly, as the histograms of the gap changes in Fig. 2f show, the average gap size over the entire field of view is preserved: the gap is both reduced and enhanced in equal measure in the vicinity of the manipulated atoms.

How does the spectral gap change with respect to the structural modifications? First we calculate the difference in peak-to-peak gap before and after field treatment, $\delta\Delta_{2\text{-}1}(\mathbf{r}) = \Delta_2(\mathbf{r}) - \Delta_1(\mathbf{r})$. The resulting image for Figs. 2c-d is shown in Fig. 3a-b. Then, by comparing the simultaneously taken topographies, we identify the locations where the surface Bi atoms have been manipulated. These locations are indicated by dots in Fig. 3a. Similarly, from high energy differential conductance map we can extract the position of all near-surface oxygen dopants (dots in Fig. 3b), as well as pinpoint which ones have changed (marked green in Fig. 3b). Clearly, the gap change is linked to the atomic manipulations, but in a rather unexpected manner: the altered sites mark the boundary between regions of increasing and decreasing gaps, whereas the gap on the sites itself is hardly affected. Interestingly, whereas one would expect the manipulation of dopants to have a strong effect on the gap, their contribution to the gap changes does not appear to be the dominant one:

the correlation between their location and where the gap changes is only moderate, and when we adjust a single near-surface dopant atom, the gaps in its vicinity shift by a few meV at most, see Fig. S5. On the other hand, the sites where Bi atom have been manipulated (Fig. 3a) show a near-perfect match to the gap-change boundaries. Upon closer inspection two observations stand out: 1) whenever a surface modification reverts in a subsequent electric field treatment, the gap reverts back as well (see Fig. 3c-d), and 2) the direction along which the gap changes appears to be linked to the direction of the surface atom repositioning. As can be seen from Figs. 3c-d and as shown schematically in Fig. 3e, each time a Bi atom sinks below the BiO plane, a neighbouring atom laterally shifts towards the void, leading to an enhancement of the spectral gap in the direction of this shift.

To quantify the direction of the gap change in more detail, we average the individual $\delta\Delta(\mathbf{r})$ images, like those in Figs. 3c-d, after aligning their direction of maximum gap increase (see Supplementary Information section IV). As Fig. 4a shows, a clear dipole profile, centred at the surface modification in an otherwise unaffected environment, is obtained. The two lobes of the dipole have opposite sign, are a few nm in size and at opposite ends 1-2 nm distance from the surface manipulation. The horizontal line-cut on the right side of Fig. 4a shows this in more detail. The most straightforward way to create a dipole-shaped profile is by subtracting two identical (2-dimensional) peaks that are laterally shifted with respect to each other, see Fig. 4b. Importantly, the maximum of the subtraction of two peaks is in the direction of their shift, as we observe in experiment. For shifts smaller than the width of the peaks, the lateral extent of the resulting dipole will depend exclusively on the width of the peaks, whereas the magnitude of the dipole is a function of the size of the shift and the amplitude of the peaks (inset Fig. 4b). Using this simple toy model, we can accurately reproduce both the size and shape of the observed dipole when we use a Gaussian profile for the peaks. Conversely, a Lorentzian profile decays too slowly to properly fit the tails of the difference image, as can be seen from the line-cut comparison in Fig. 4a. For a realistic lateral shift of 2 Å (see e.g. Fig. 3c) the Gaussians require an amplitude of ~50 meV, which is not unreasonable given the average peak-to-peak gap size of $\Delta$ = 96 meV. Lastly, the width of the Gaussians is 1.7 nm, which is comparable to the superconducting coherence length.

A Gaussian shaped gap-distribution that falls off exponentially with length scales on the order of the coherence length is highly suggestive of a point-like object that creates a local pairing potential. The physical origin of this object is likely related to oxygen dopants and/or apical oxygen atoms. Previous studies found a correlation between the strong spatial variation of the spectral gap and the oxygen dopant atoms (*8*, *9*). However, the correlation for the near-surface dopants we manipulate (i.e. those with a resonance at E ~ -1 eV) was shown to be relatively small compared to deeper lying ones (*9*), suggesting their contribution to the total gap is limited. The modest influence on the gap we observe upon manipulating these dopants is in agreement with this notion. Furthermore, the large gap changes, that have a one-to-one correspondence with the Bi atom surface modifications, do not necessarily have a dopant atom nearby. On the other hand, the shift of the surface Bi atoms is plausibly concomitant with a shift of the apical oxygen atoms that are directly below them. The importance of the apical oxygen atoms in both the tunnelling process (*23-25*) and the gap size (*10*) have been stressed previously. Our observations provide additional input for further theoretical studies, in particular those that take into account on-site correlations (*26*, *27*), into the origin of the spectral gap, and the tunnelling process, in this cuprate superconductor.

An alternative source of the dipole-shaped difference image is the appearance of topological defects that introduce a $2\pi$-rotation of the phase of the order parameter (*28*). However, creation and annihilation of topological defects has to occur in pairs. Given our finite field of view, it may be that one half of the pair is outside our measurement range, but based on the large field of view topographic before-after comparison (Fig. S4), and assuming that both pairs will have a signature in topography, this is unlikely. Additionally, in our optimally doped system we do not find a significant correlation between the gap changes and topological defects in the smectic, or the d-form factor density wave, reported for underdoped samples (*29*, 30). Extension of our work to lower doping concentrations, where the various charge- and spin ordered states are more predominant, should give a more definite answer to this issue. Additionally, the influence of dopants and the surface structure on these (ordered) states themselves can be studied directly using the field-induced atom manipulation we introduce in this work.

The observation of a profound influence on the peak-to-peak gap in tunnelling experiments of sub-nanometre shifts in atomic positions highlights the importance of the lattice on the local electronic properties of the cuprates. The spatial profile of the gap modification we observe is highly suggestive of the field induced lateral movement of point-like objects that create a local pairing potential in the $CuO_2$ plane. This work demonstrates a new avenue to non-invasively and reversibly probe the influence of the local lattice on the electronic states of cuprate high temperature superconductors and related compounds.

**Acknowledgments:** We thank M. Civelli, A. Mesaros and P. Simon for fruitful discussions.

**Funding:** FM would like to acknowledge funding from H2020 Marie Skłodowska-Curie Actions (grant number 659247) and the ANR (ANR-16-ACHN-0018-01).

**Supplementary Materials:**

Materials and Methods

Figures S1-S9

References (*31-35*)

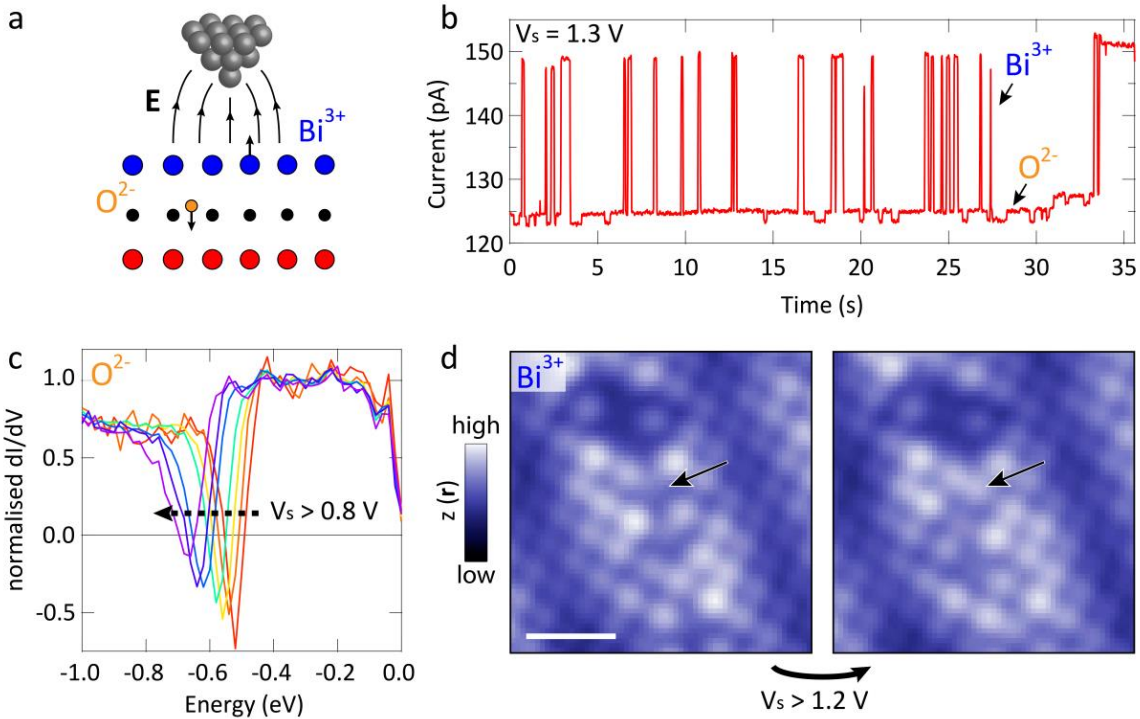

**Fig. 1.** Electric field induced atom manipulation. (**A**) Schematic representation of the field induced manipulation of near surface oxygen dopants and surface Bi atoms (blue = Bi, black = O, red = Cu). (**B**) Current as function of time at $V_s = 1.3$ V: small jumps signal dopant manipulation, large jumps surface Bi rearrangement. (**C**) Differential conductance on the same location after sequential $V_s > 0.8$ V treatment at a current of ~100 pA ($E_{set} = -1$ eV, $I_{set} = 400$ pA). (**D**) Topography before (left) and after (right) $V_s > 1.2$ V treatment: surface Bi atoms on the crest of the super-modulation are manipulated, one instance is indicated by arrows ($E_{set} = -100$ meV, $I_{set} = 100$ pA).

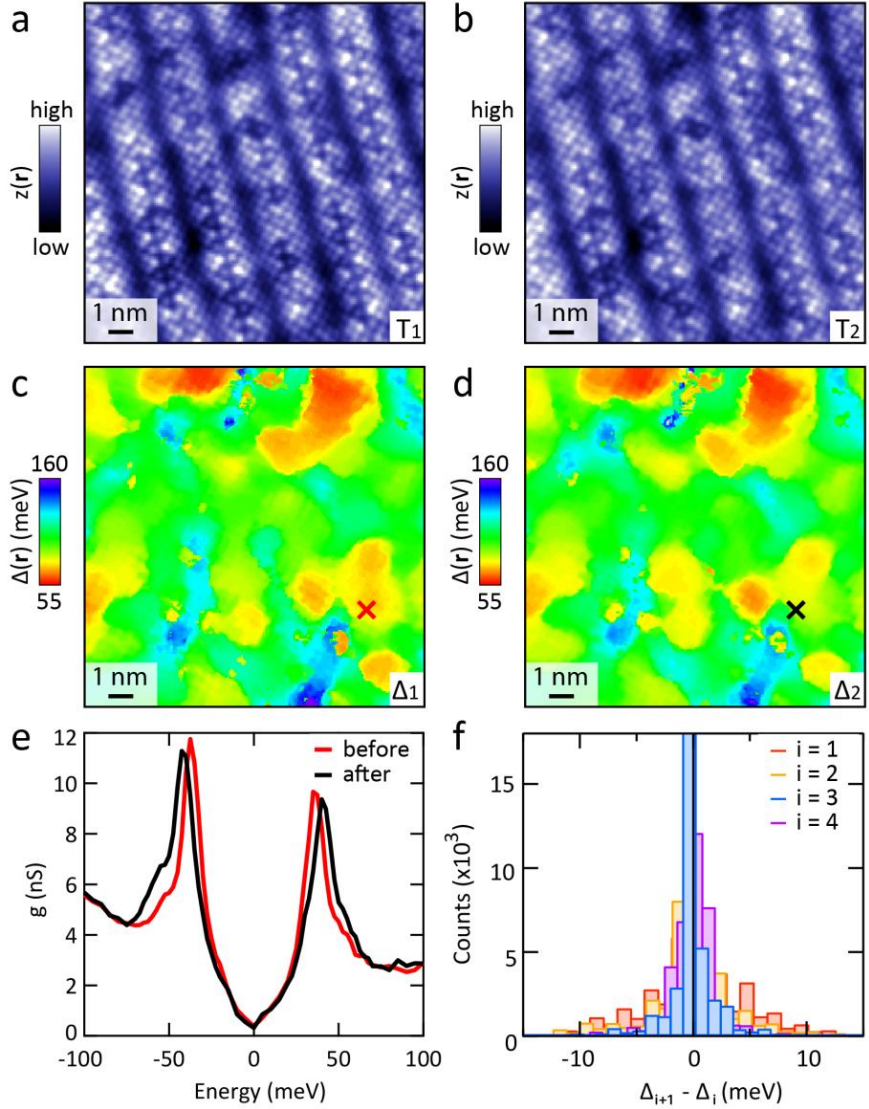

**Fig. 2.** Gap modification. Constant current image taken before (**A**) and after (**B**) field induced atom manipulation at $V_s = 1.5$ V ($E_{set} = 100$ meV, $I_{set} = 400$ pA). (**C, D**) Maps of the peak-to-peak gap, $\Delta$, corresponding to the topographies in (A, B). (**E**) Spectrum taken before and after manipulation on the location indicated by a cross in panels (C, D) showing a clear change in $\Delta$. (**F**) Histograms of $\delta\Delta = \Delta_{i+1} - \Delta_i$, for five subsequent maps, i=1 corresponds to the maps in (C, D): for all maps the net change in the gap size is zero.

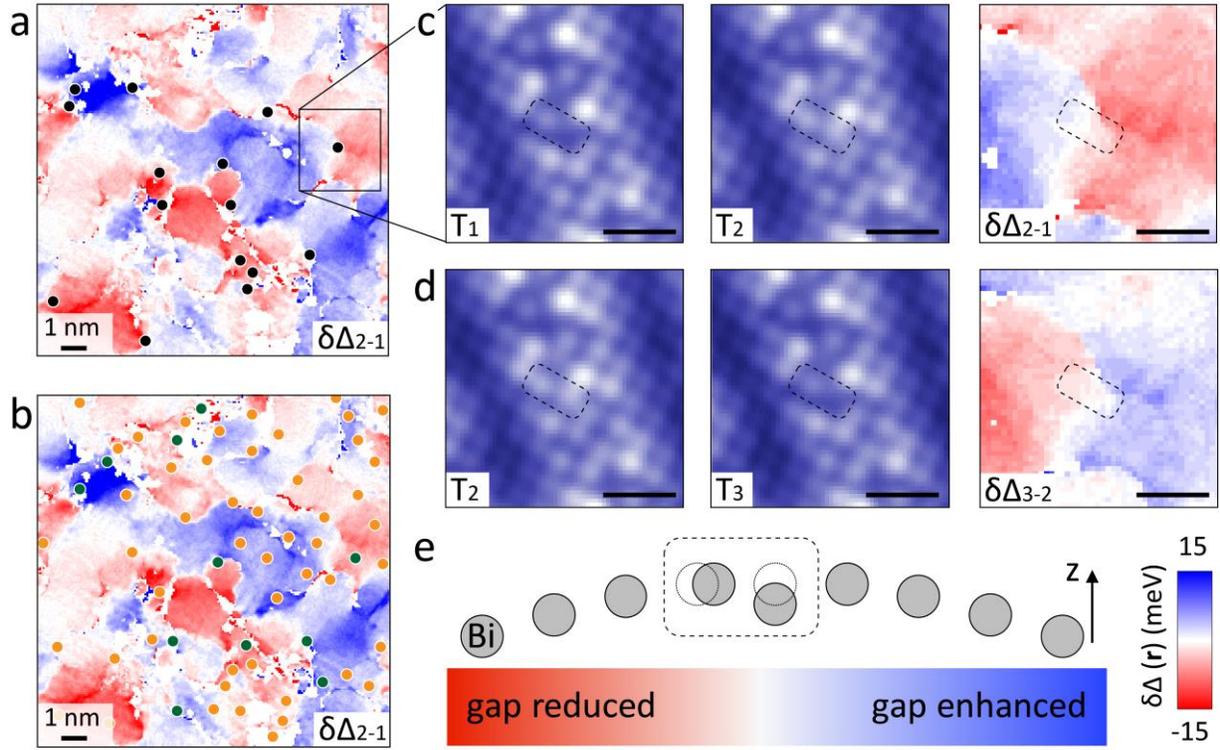

**Fig. 3.** Atomic position dependence of gap changes. (**A**) Map of the difference in the gap, $\delta\Delta_{2\text{-}1}(\mathbf{r}) = \Delta_2(\mathbf{r}) - \Delta_1(\mathbf{r})$. All locations where Bi atoms have been manipulated are indicated by a black dot. (**B**) Same as (A), but here all near-surface dopants are indicated with dots, manipulated dopants are in green. (**C**) Enlargement of the topography before (left) and after (middle) manipulation in the area marked by a box in (A) (colour scale as in Fig. 2a). The manipulated Bi atoms are highlighted by the dotted box, the corresponding $\delta\Delta_{2\text{-}1}(\mathbf{r})$ is shown in the right panel. (**D**) As (C), but after a subsequent $V_s = 1.5$ V manipulation reverts the Bi atoms and the gap back to their original configuration. The scale bars in (C-D) indicate 1 nm. (**E**) Schematic of the surface change and its effect on the gap: one atom moves down, another shifts laterally towards it resulting in gap enhancement in the direction of the shift. The labelling of the images is analogous to Fig. 2a-d. More details and examples can be found in SI section I and IV.

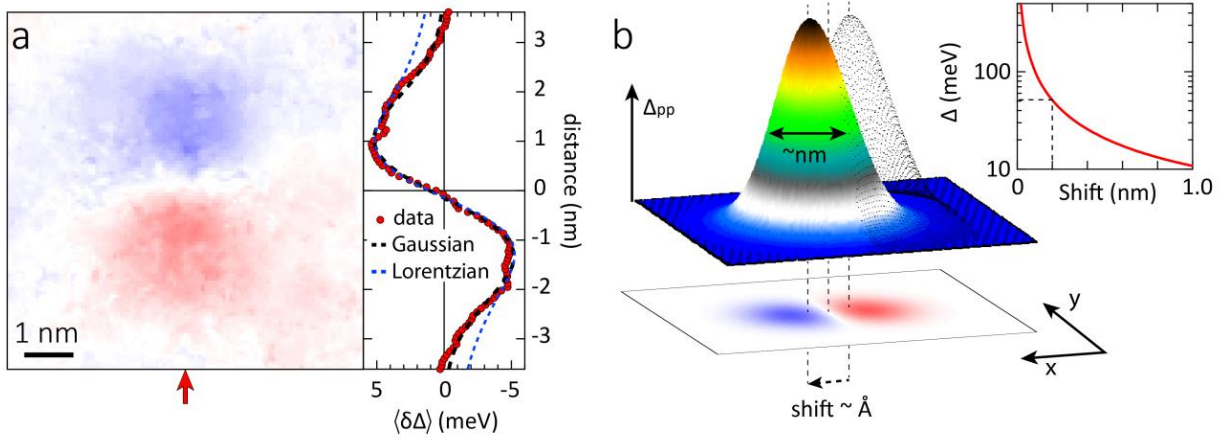

**Fig. 4.** Dipole profile and toy model. (**A**) Average $\delta\Delta(\mathbf{r})$ for all topographic modifications after aligning their orientation. The arrow indicates where the vertical line trace on the right is taken for the experimental data (red), and for the toy model using a Gaussian (black) and Lorentzian (blue) peak profile. (**B**) 2D-Gaussian (width = 1.7 nm) that is shifted by a fraction of a nm (top, shift exaggerated for clarity), leading to a difference plot (bottom) with the same shape and length-scales as the experimental data in (A). The inset shows how the amplitude of the subtracted Gaussians depends on the shift distance: a 2 Å shift requires an amplitude of $\Delta \sim 50$ meV (dashed lines). Colour scales as in Fig. 3.